# DSCMC: Distributed Stateless Code Model Checker

Elaheh Ghassabani and Mohammad Abdollahi Azgomi[*]

Trustworthy Computing Laboratory, School of Computer Engineering,
Iran University of Science and Technology, Tehran, Iran

E-mail: ghasabani@comp.iust.ac.ir, azgomi@iust.ac.ir

**Abstract**

*Stateless code model checking* is an effective verification technique, which is more applicable than stateful model checking to the software world. Existing stateless model checkers support the verification of neither LTL formulae nor the information flow security properties. This paper proposes a *distributed stateless code model checker* (DSCMC) designed based on the Actor model, and has the capability of verifying code written in different programming languages. This tool is implemented using *Erlang*, which is an actor-based programming language. DSCMC is able to detect deadlocks, livelocks, and data races automatically. In addition, the tool can verify information flow security and the properties specified in LTL. Thanks to its actor-based architecture, DSCMC provides a wide range of capabilities. The parallel architecture of the tool exploiting the rich concurrency model of Erlang is suited to the time-intensive process of stateless code model checking.

## 1. Introduction

Although concurrent programming has become more prevalent in recent years, it is so notoriously error prone that traditional testing methods (such as various forms of stress and random testing) are inadequate to detect all concurrency errors existing in a program [1]. *Model checking* [2] is a promising method for detecting and debugging concurrency errors [1, 2, 3], but *traditional model checkers* are rather appropriate to verify code written in general purpose programming languages because to use such tools, users have to manually model their programs as input to a model checker. So the validity of the verification result relies on that model, hence it is necessary that the input model conform to the target code. However, modeling needs special knowledge and expertise [4]. A more appropriate approach in the software world is to use *code model checking*, where model checkers are able to verify realistic code instead of a model.

Code model checking can be divided into different branches. From one point of view, it can be categorized as stateful and stateless model checking. Theoretically, stateful techniques are ideally suited to verify concurrent programs, but, in practice, they usually run into state space explosion during verifying parallel programs. A stateful model checker has to capture all the essential states of the program so when the concurrency level rises, the state space grows exponentially, which may be out of control (we may face state space explosion). In such situations, stateless model checking can be useful. This technique is especially appropriate to explore the state space of large programs because it is indeed arduous to precisely capture and control all the essential states of a large program [1, 5]. The state of a running program includes global variables, heap, thread stacks, and register contexts, and so on. Even if the whole state space could be captured and controlled, processing such large states would be very expensive [6, 7].

---
[*] Corresponding author. Address: School of Computer Engineering, Iran University of Science and Technology, Hengam St., Resalat Sq., Tehran, Iran, Postal Code: 16846-13114, Fax: +98-21-73021480, E-mail: azgomi@iust.ac.ir.

The idea of stateless model checking was proposed by Godefroid, simultaneously with the appearance of code model checking [5]. A stateless model checker explores the program state space without requiring to capture any program states. The program is executed under the control of the scheduler of the stateless model checker. This special scheduler systematically explores all the execution paths of the program obtained by nondeterministic thread interleavings [1, 8]. In this paper we develop a new stateless model checker called DSCMC (distributed stateless code model checker). In summary, the main novel contributions of the tool can be pointed out as follows:

- DSCMC is written in Erlang, hence it exploits Erlang's strong capabilities, including integrating different programming languages. By employing this feature, our tool is able to verify any program written in an arbitrary language if that program has been instrumented according to DSCMC's instrumentation rules.
- Stateless model checking is a time-intensive process with the runtime checking overhead. A parallel and distributed tool can compensate for this drawback. DSCMC is a parallel and distributed stateless model checker developed based on the *Actor model*.
- As far as we know, existing stateless model checkers do not support verifying linear temporal logic (LTL) formulae; we design a new method for stateless model checking of the LTL formulae in DSCMC. This method exploits the actor-based nature of DSCMC.
- DSCMC is the first stateless model checker that is able to check information flow security in concurrent programs using a new method. Moreover, DSCMC has other capabilities suited to verify concurrent programs, including detecting assertion violation, deadlocks, livelocks, and data races.

The remainder of this paper has the following structure. Section 2 gives some required background. Section 3 elaborates the design of the proposed tool. Section 4 describes the implementation issues and the code instrumentation process. Section 5 describes the method applied to verify the tool. Section 6 covers the related work. Finally, Section 7 mentions some concluding remarks.

## 2. Preliminaries

This section presents required background for this paper. The first subsection introduces the actor model briefly. The second subsection gives the required notions and definitions for the paper.

## 2.1 Actor model

*Actor* is a model for concurrent computing to develop parallel and distributed systems. Each actor is an autonomous entity that acts asynchronously and concurrently with other actors. An actor can send/receive messages to/from other actors, create new actors, and update its own local state. An actor system is composed of a collection of actors, some of whom may send messages to (or receive messages from) actors outside the system [9]. Therefore, the actor primitives are *send*, *become*, *newadr*, and *initbeh* [10]. An actor via command *send(a, v)* creates a new message with receiver *a* and contents *v*, and puts it to the message delivery system. This system guarantees the received message will be finally delivered to actor *a*. Command *become(b)* creates a new anonymous actor to perform the remains of the current computation, alters the behavior of the actor executing *become* to be *b*, and frees that actor to accept another message. The anonymous actor primitives are the same as other actors, but it will never receive any messages because its address is unknown. *newadr* and *initbeh* are used for creating actors. Command *newadr()* creates a new actor and returns its address. *initbeh(a, b)* initializes the behavior of a newly created actor with address *a* to be *b*. The behavior of actor *a* can only be initialized by the actor who created *a*. Actor's behavior is described by lambda abstraction, which embodies the code that should be executed after receiving a message [10].

We use *Rebeca modeling language* [11, 12] to design and verify DSCMC. Rebeca is an actor-based modeling language with a formal foundation [13, 14]. We have aimed to develop a parallel code model checker so we decided to design the tool by using the Actor model. Therefore, our tool has been modeled in Rebeca based on Actors, and implemented using Erlang. Erlang is arguably the best known implementation of the Actor model [9]. All the components of DSCMC have been modeled in Rebeca.

Rebeca is a Java like language, in which a model consists of concurrently executing reactive objects, *rebecs*. The *rebecs* carry out the computation by asynchronous message passing and executing the corresponding methods of messages. Each message is put in the unbounded queue of the receiver *rebec*. For servicing a message, a specified method should be invoked [14].

An actor language is an extension of a functional language. The algorithms presented in this paper are written in Erlang syntax. In the Erlang language, each actor corresponds to a process. The process primitives are *!* (i.e., *send*), *receive*, and *spawn* (i.e., creating a new process). For more information, please see [15, 16, 17].

Erlang is a functional language; therefore everything in this language is a function. By using Erlang functions, programmers are able to write both sequential and parallel programs. When a process is created by *spawn*, an Erlang function is specified as the behavior of this process. A *function declaration* is a sequence of function clauses separated by semicolons, and ended with period (.). A *function clause* has a clause head and a clause body that are separated by ->. A clause *head* contains the name of the function, a list of input argument, and an optional guard sequence beginning with keyword *when*. A clause *body* consists of a sequence of valid expressions separated by comma (,) [16]. Valid Erlang expressions and guard sequences are described in [16, 17].

Command *spawn (Module, Function, Args)* in Erlang corresponds to *newadr* and *initbeh* primitives in the Actor model. This command creates a new process (actor) that behaves corresponding to function *Function* where *Args* is a *list* of its arguments, and *Module* is the name of the module where *Function* is defined. After creating a new process, *spawn* returns the address of the newly created process (i.e., *Pid*). A process is able to create a new process at every point of its behavioral function. The newly created process carries out the rest of the computation (it is similar to *become* in the Actor model if the return *Pid* is not saved). The command of the form *Pid ! Msg* sends message *Msg* to the process with address *Pid*. In Erlang, command *receive* has the following syntax:

```
 receive
     Pattern1 when Guard1  -> exp11, .., exp1n;
     Pattern2 when Guard2  -> exp21, .., exp2n;
     ...
     Other    -> expn1, .., expnn
end
```

When a process receives a message, it tries matching that message against *Pattern1* (with respect to the possible guard, *Guard1*); if the pattern matching succeeds, it evaluates *Expressions1*, or else it tries *Pattern2*, and so on. If none of the patterns matches, the message is saved for later processing, and the process waits for the next message. The patterns and guards used in a *receive* statement have exactly the same meaning and syntax as the patterns and guards used for defining a function [16]. In order to get familiar more with Erlang, please see [16, 17].

**2.2 Definitions**

A concurrent system is composed of a finite set of threads or processes whose state space is defined using dynamic semantics in the style of [2]. Each process executes a sequence of statements in a deterministic sequential programming language, such as C, C++ or Java. Threads are thus a particular type of processes that share the same heap [18]. A multi-threaded program can be modeled as a concurrent system, which consists of a finite set of threads and a set of shared objects. Threads communicate with each other only through shared objects. Operations on shared objects are called visible operations, while the rest are invisible operations. A state of a multi-threaded program contains the global state of all shared objects and the local state of each thread. In a multi-threaded program, a visible operation performed by a thread is considered as a transition that advances the program from one global state to a subsequent global state. Such a transition is followed by a finite sequence of invisible operations of the same thread, ending just before the next visible operation of that thread [19].

To avoid exploring redundant interleavings, we use dynamic partial order reduction (DPOR) [18] because the number of possible interleavings grows exponentially as the program is getting large. DSCMC also uses the algorithm of DPOR described in [20].

Partial order reduction algorithms only explore a proper subset of the enabled transitions at a given state *s* such that it is guaranteed to preserve the interested properties. DPOR dynamically tracks threads interactions to identify points where alternative paths in the state space need to be explored [18, 20].

Stateless model checkers explore the program state space by executing the program in its realistic runtime environment and observing its visible operations. A stateless model checker does not keep the search history because storing and restoring the state of a program that concretely runs is so expensive [20]. The program state space is explored by repeating the process of the program execution under different scheduling choices. In order to verify program *P*, each iteration of the stateless model checking (the *SMC* function) is equivalent to execute *P* under a special scheduler.

Given a state *s* and a transition *t*, we use some notations (from [20]); *s.enabled* denotes the set of transitions that are enabled from s. A thread *p* is enabled in a state *s* if there exists some transition *t* such that $t \in s.enabled$ and *t.tid = p* where *t.tid* refers to the identity of the thread that executes *t*. *s.backtrack* denotes the backtracking set [18] at state *s*. *s.backtrack* is a set of thread identities. Here, *{t | t.tid ∈ s.backtrack}* is the set of transitions that are enabled but have not been executed from *s*.

Assume *t, t'* ∈ *s.enabled* are two transitions enabled in state *s*, and *t* is the transition chosen to execute by the model checker. We say *t* and *t'* are co-enabled. For each to-be-executed transition *t'*, DPOR dynamically checks whether *t* and *t'* are dependent. If so *t'.tid* is added to *s.backtrack*. Later, during backtracking (in other iterations of *SMC*), if a state *s* is found with non-empty *s.backtrack*, DPOR will pick one transition *t'* such that *t'* ∈ *s.enabled* and *t'.tid* ∈ *s.backtrack*, and explore a new path of the state space by executing *t'* [20].

## 3. The DSCMC tool

This section describes design of a new code model checker using stateless model checking, called DSCMC (Distributed Stateless Code Model Checker). DSCMC is designed based on the Actor model [9, 10], and implemented using Erlang [16, 21]. DSCMC is a distributed tool; Fig. 1 shows the components of DSCMC and their interactions on each computational node.

The process of model checking in DSCMC can be summarized in the following steps:
- Instrumenting the code of the program under test.
- Running the first iteration of *SMC* on one computational node.
- Distributing detected backtracking points among available computational nodes.
- Each node continues the process of stateless model checking for the backtracking points delegated to it until it explores all the backtracking points that appear in the process.

Each component in Fig. 1 is an actor (or a collection of actors) in DSCMC. Naturally, these actors should interact with one another by message passing. DSCMC's actors are mapped to Erlang processes. It is worth pointing out some features of Erlang processes; Erlang processes are very lightweight and cheap to create (about 100 times lighter than threads [22]). The Erlang message passing mechanism is also very fast [16] (about one micro second [16, 22]). Therefore, there is no concern about both process creation and the message passing overhead. Due to using the Actor model, DSCMC has a modular architecture which can help to easily promote its algorithms and components individually as well as extend its functionality.

The architecture of DSCMC makes it possible to verify programs written in various programming languages. For this reason, DSCMC uses two interfaces: (1) *program side interface (Pi)*, and (2) *tool side interface (Ti)*. *Pi* is a library with functions that are used for code instrumentation. This library can be implemented for any arbitrary programming language in the same way as different programming languages are integrated with Erlang, said in [16, 17]. *Pi* is also a controller at the program side, and regulates communications between the program and DSCMC's actors. In fact, this interface is an actor who formats the required part of the program state as a message, and sends it to actor *Ti*. All requests for communications with the scheduler and other DSCMC's actors are sent by *Pi* to *Ti*. *Ti* sends received messages, based on their headers and contents, to DSCMC's actors.

To use the DSCMC tool, first, code must be instrumented. The code is instrumented such that the state of the executing program can be monitored and the program can communicate with DSCMC. Thereafter, the program is compiled into an executable and DSCMC runs the executable repeatedly under its scheduler until all relevant interleavings among the threads have been explored. Before performing any operation that might have side effects on the other threads, the instrumented program sends a request to the scheduler. The scheduler can block the requester by postponing a reply. *Pi* controls the execution of the multi-threaded programs on a multi-core processor such that the execution on a single-core processor can be emulated; because the interleaving notion assumes that there is a scheduler which interlocks the steps of concurrently executing threads without any known strategies and thus models any possible realization by a single-processor machine or by several processors with arbitrary speeds [2]. At runtime, threads yield the processor at specific execution points using *Pi* (these points are the same preemption points described in Section IV). When the current executing thread reaches the end of its time slice, it yields the processor to *Pi*. *Pi* informs DSCMC's scheduler about the current status of threads so that DSCMC's scheduler can give the processor to an enabled thread upon the fair scheduling algorithm [1].

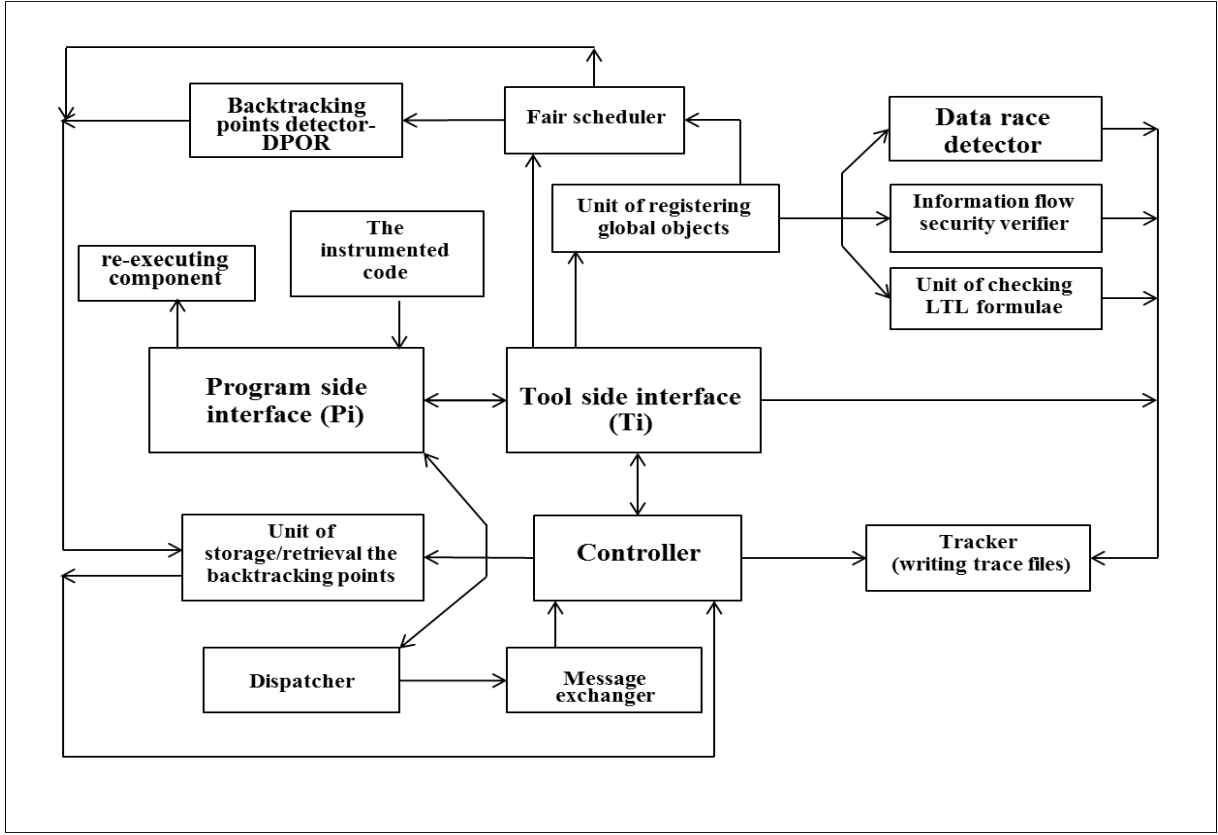

**Fig. 1.** The architecture of DSCMC

Fig. 2 shows the process of stateless model checking in DSCMC. This process is modeled in Rebeca as a parallel process in Fig. 3, which shows the main actors of DSCMC and their interactions. To construct an abstract model, some actors have been omitted from this model, including actors *Data race detector*, *Information flow security verifier*, and *Unit of checking LTL formulae*. Note that stateless model checking is a dynamic exploration technique whose primary algorithms are DPOR and scheduling. In DSCMC, the scheduling algorithm is the same fair scheduling algorithm [1] and the algorithm used for DPOR is the same as [18].

Distribution of the tool has no effect on these internal algorithms. In DSCMC, backtracking points are distributed in a different way from [23]. DSCMC initiates the *SMC* function from one computational node. During the first iteration, some backtracking points in the state space are detected, and thereafter DSCMC's *dispatcher* distributes these backtracking points among the computational nodes with respect to the number of both available nodes and detected backtracking points. Each node may visit many new backtracking points while exploring backtracking points delegated to it. Therefore, after the first iteration of stateless model checking, each node explores backtracking points delegated to it, and visited by it until it explores all of them. When a node explores all its backtracking points, it ends the process of model checking. When all nodes end the model checking process, the whole process of stateless model checking terminates (in the further work, it is possible to add a load balancer to DSCMC). By employing a fair scheduler and this method for distributing backtracking points, the nodes explore different backtracking points. The idea of fair stateless model checking and its benefits have been described in [1]. We also use this idea for state space exploration on each node. An unfair schedule corresponds to a cycle in the state space of a correct program, in which an enabled thread is starved continuously. By exploring the state space using a fair scheduler, the model checker is able to prune such cycles away. Note that a cycle in an incorrect program might correspond to a livelock [1].

Erroneous cycles in the state space must be fair otherwise they would not be considered a programming error. DSCMC's scheduler prune unfair cycles and will generate an infinite execution in the limit for fair cycles. Fair stateless model checking has the ability to detect livelocks by distinguishing between fair and unfair executions [1].

Now, let us describe the process of stateless model checking in DSCMC by using the model in Fig. 3. The process is initiated from *tool side interface* (i.e., actor *mci* in Fig. 3, line 66). The *mci* actor creates other DSCMC's actors, and then creates the *program side interface* actor (actor *pi* in Fig. 3, line 67). After creating *pi*, the executable of the instrumented program will be executed (line 50), and its execution will be monitored by *pi*. At runtime, *pi* sends the required status of the program to *mci* in the form of a proper message. Whenever *mci* receives a message pertaining to the scheduler, it sends the message to the scheduler (i.e., actor *sch* at line 69). After receiving the message, the scheduler should choose an enabled thread according to the scheduling algorithm in [1] (such messages give information about the status of the threads in the executing program to the scheduler). Thereafter, the identity of the selected thread is sent to *pi* by the scheduler at line 93. Among previous stateless model checkers, only the CHESS is able to detect livelocks. As we mentioned before, we also follow the CHESS approach to detect livelocks. During state space exploration, our scheduler selects an enabled thread; if there is no enabled thread when the current iteration has not ended yet, a deadlock has occurred. Therefore, the scheduler reports the deadlock situation to the controller, and then the controller terminates the current iteration.

When the program reaches an undesired state during state space exploration by DSCMC, the process of model checking does not end, and only current iteration that caused the violated path is terminated. DSCMC saves the counterexample (the violated path) for further inspection, and continues exploring the state space for another execution path (this is one of the strengths of DSCMC). The counterexample indicates an execution path from the initial state to the violated state. By employing the *re-executing* and *tracker* components in Fig. 1, the user can replay a violated scenario, obtain useful debugging information, and adapt the program (or the property) accordingly. The component of DSCMC that stores trace files is the same actor *tracker* in Fig. 1 and Fig. 3. Trace files are used during backtracking and re-executing a counterexample. A trace file contains information about how threads were scheduled in a special path. By using a trace file, its related path is re-executed. In order to construct a trace file, the scheduler should write the identity of the selected thread to the current trace file. Therefore, after scheduling a thread, the scheduler sends the identity of the selected thread to *tracker* (Fig. 3, line 95). When a violation, including deadlocks, livelocks, and data races, occurs, the tracker writes the violated path in a separate trace file with a proper name so that the user can identifies the file (lines 97, 98). During state space exploration, if DSCMC reaches a new backtracking point, it saves required information about that point for further exploration (line 114). This information is needed for fair scheduling in the backtracking phase.

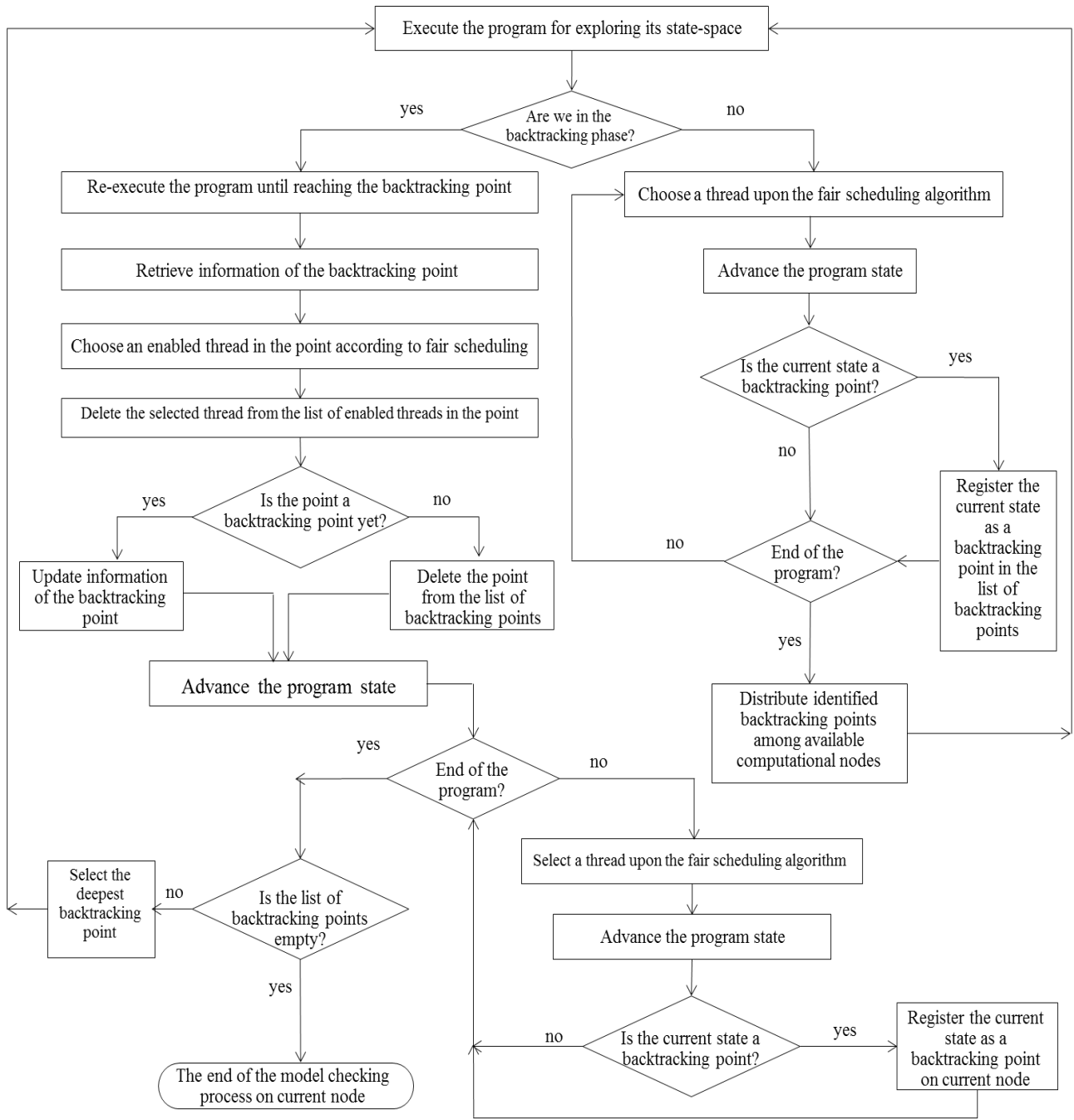

**Fig. 2.** The process of stateless model checking in DSCMC

The controller, shown in Fig. 1, is the same actor *cntrl* in Fig. 3. When an execution path terminates (i.e., an iteration of *SMC*), the scheduler informs the controller that it should start to explore another one (line 24). The controller also informs other actors about ending one execution path (Fig. 3, lines 26, 31, and 37). Receiving such a message from the controller, each actor shows a proper behavior. In this phase, the controller must select a backtracking point for the next exploration. Various strategies can be taken to select this point. We have the strategy as follows. After ending a path, the controller selects the deepest backtracking point for the next exploration (lines 21-23). The deepest backtracking point is the latest *detected* point in previous iterations. Selecting the deepest point, its information is retrieved and *mci* restarts the program, and replays it from the initial state until reaching the selected backtracking point (lines 74-76).

When the program gets to the selected backtracking point, the scheduler chooses a thread that is enabled but has not been scheduled in this state (line 77). Thereafter, this backtracking point omitted from the list of backtracking points at line 105. If the current backtracking point still contains any enabled thread from which has not been scheduled, the information about the point is updated; i.e. the thread selected from the point is omitted from the list of enabled threads on the backtracking point, and then the DPOR algorithm is applied on this state (line 106), then if the state is still a backtracking point (after applying the algorithm), its information is updated at line 126.

The task of actor *dpor* (Fig. 3) is to apply the algorithm DPOR. More details about this algorithm have been described in [18, 25]. In each state, actor *dpor* checks whether the state is a backtracking point. That is, in the current state, if there are at least two enabled and dependent transitions which have not been scheduled yet, the state may be a backtracking point. After applying the DPOR algorithm on that state, if there are at least two dependent threads, the state is considered as backtracking point. If the current state is a backtracking point, its information is captured by actor *btset* (lines 126 and 133). In the first prototype of DSCMC this information is stored with the Erlang DETS mechanism [16, 21].

After ending the first iteration of stateless model checking, the function whose task is to distribute the workload is called (lines 142-145). Sending and receiving information is done with the *message exchange* component in Fig. 1 (Fig. 3, lines 151-152). When a node receives the workload, its controller starts to explore the deepest backtracking point in its workload list. Fig. 3 shows DSCMC's actors for one node. The actors on the other nodes behave in the same way except that actor *disp* ends its duty after the first iteration of *SMC*. The process of stateless model checking on the other nodes is the same as the second iteration (or further iterations) on the first node. When all the backtracking points on one node are explored, the controller terminates the process of model checking on the node (line 22).

```
1. reactiveclass Controller(5) {
2.   knownrebecs {
3.     Dispatcher disp;  Tracker tracker;
4.     Bt_point_rw btset; Prog_side_interface pi;
5.     Mc_side_interface mci;
6.   }
7.   statevars {
8.   boolean backtrack_phase;  boolean mc;
9.   int iteration; int bt;
10.  }
11.  msgsrv initial( ){
12.    backtrack_phase = false;
13.    iteration = 0;  bt = 1;
14.  }
15.  msgsrv prep_next_iteration( ){
16.    if( !backtrack_phase ) {
17.      backtrack_phase = true;
18.      disp.dispatch( );
19.    }
20.    iteration = iteration + 1;
21.    mc = ? (true, false);
22.    if (mc) { //end of MC on local node }
23.    else { // bt := deepest backtrack point
24.        mci.backtrack(bt);  pi.run( ); }
25.  }
26.  msgsrv normal_end ( ){
27.    tracker.close_file(iteration);
28.    btset.close( );
29.    self.prep_next_iteration( );
30.  }
31.  msgsrv livelock_end ( ){
32.    tracker.save_livelock_trace(iteration);
33.    tracker.close_file(iteration);
34.    btset.close( );
35.    self.prep_next_iteration( );
36.  }
37.  msgsrv deadlock_end ( ){
38.    tracker.save_deadlock_trace(iteration);
39.    tracker.close_file(iteration);
40.    btset.close( );
41.    self.prep_next_iteration( );
42.  }
43.  msgsrv reaching_to_bt( ){ btset.load(bt); }
44. }
45. reactiveclass Prog_side_interface(5) {
46.  knownrebecs { Mc_side_interface mci; }
47.  statevars { int prog_msg; }
48.  msgsrv initial( ){ prog_msg = 0; }
49.  msgsrv run( ) {
50.    self.program( );
51.    self.rcv_from_sch(0);
52.  }
53.  msgsrv program( ){
54.    prog_msg = ? (0, 1, 2);
55.    self.send (prog_msg);
56.  }
57.  msgsrv send(int msg){ mci.from_prog(msg); }
58.  msgsrv rcv_from_sch(int thread){ self.program( ); }
59. }
60. reactiveclass Mc_side_interface(5) {
61.  knownrebecs {
62.    Prog_side_interface pi; Controller cntrl;
63.    Fair_scheduler sch;   Sh_obj_reg oid;
64.  }
65.  statevars { }
66.  msgsrv initial( ){ self.start_mc( ); }
67.  msgsrv start_mc( ) {   pi.run( );  }
68.  msgsrv from_prog(int msg) {
69.    if (msg == 0)   { sch.thread_st(msg); }
70.    else if (msg == 1){ oid. thread_st(msg); }
71.    else if (msg == 2){ oid. register(msg); }
72.  }
73.  msgsrv backtrack(int bt) {
74.    // backtracking from trace file of bt
75.    // reaching to bt
76.    cntrl.reaching_to_bt( );
77.    sch.sch_bt(bt);
78.  }
79. }
80. reactiveclass Fair_scheduler(5) {
81.  knownrebecs {
82.    Controller cntrl; Prog_side_interface pi;
83.    Tracker tracker; Dpor dpor; Bt_point_rw btset;
84.  }
85.  statevars {
86.    int tid;  byte status;
87.  }
88.  msgsrv initial( ){  }
89.  msgsrv thread_st(int msg) {
90.    if(msg == 0){  cntrl.normal_end( ); }
91.    else{
92.      tid = ? (0,1,2,3,4);
93.      pi.rcv_from_sch(tid);
94.      dpor.chk(msg);
95.      tracker.save(tid);
96.      status = ?(0,1,2);
97.      if (status == 0){ cntrl.deadlock_end( ); }
98.      if (status == 1){ cntrl.livelock_end( ); }
99.    }
100. }
101. msgsrv sch_bt(int bt) {
102.   // tid := one_of_enabled_threads in bt
103.   pi.rcv_from_sch(tid);
104.   status = ?(0,1);
105.   btset.del(bt);
106.   if(status == 0}{ dpor.chk(bt-tid); }
107. }
108. }
109. reactiveclass Tracker(5) {
110.  knownrebecs { }   statevars { }
111.  msgsrv initial( ){ }
112.  msgsrv save(int tid) { /* do something */ }
113.  msgsrv save_deadlock_trace(int name)
114.  { /* do something */ }
115.  msgsrv save_livelock_trace(int name)
116.  { /* do something */ }
117.  msgsrv close_file(int name){/* do something */}
118. }
```

**Fig. 3.** Modeling DSCMC's actors and their interactions in Rebeca

```
119. reactiveclass Dpor(5) {
120.   knownrebecs { Bt_point_rw btset; }
121.   statevars { boolean is_bt_point; }
122.   msgsrv initial( ){ }
123.   msgsrv chk (int msg) {
124.     // DPOR algorithm
125.     is_bt_point = ?(true, false);
126.     if (is_bt_point) { btset.save (msg); }
127.   }
128. }
129. reactiveclass Bt_point_rw(5) {
130.   knownrebecs { Controller cntrl; }
131.   statevars {  }
132.   msgsrv initial( ){ /* do something */ }
133.   msgsrv save(int msg) { /* do something */ }
134.   msgsrv load(int bt){ /* load information */ }
135.   msgsrv del(int bt){ /* delete the bt */ }
136.   msgsrv close( ) { /* close DETS */  }
137. }
138. reactiveclass Dispatcher(5) {
139.   knownrebecs { Send_rcv_workload trans; }
140.   statevars { boolean file; }
141.   msgsrv initial( ){ }
142.   msgsrv dispatch( ) {
143.     file = true;
144.     trans.send(file);
145.   }
146. }
147. reactiveclass Send_rcv_workload(5) {
148.   knownrebecs { Controller cntrl; }
149.   statevars { }
150.   msgsrv initial( ){ }
151.   msgsrv send (boolean file) { /* send data */ }
152.   msgsrv rcv (boolean file) {
153.     // receiving data from master node
154.     // This function operates on worker nodes.
155.   }
156. }
157. reactiveclass Sh_obj_reg(5) {
158.   knownrebecs { Fair_scheduler sch; }
159.   statevars { int converted; }
160.   msgsrv initial( ){ }
161.   msgsrv thread_st (int msg) {
162.     // preparing msg for sending to the scheduler
163.     converted = msg - 1;
164.     sch.thread_st(converted);
165.   }
166.   msgsrv register(int msg){ /* do something */ }
167. }
168. main {
169.   Fair_scheduler sch
           (cntrl, pi, tracker, dpor, btset):( );
170.   Sh_obj_reg oid (sch):( );
171.   Tracker tracker ( ):( );
172.   Dispatcher disp (trans):( );
173.   Dpor dpor (btset):( );
174.   Bt_point_rw btset (cntrl):( );
175.   Prog_side_interface pi (mci):( );
176.   Controller cntrl
           (disp, tracker, btset, pi, mci):( );
177.   Send_rcv_workload trans (cntrl):( );
178.   Mc_side_interface mci (pi, cntrl, sch, oid):( );
179. }
```

**Fig. 3.** Modeling DSCMC's actors and their interactions in Rebeca- *continuation*

The *unit of registering global objects* in Fig. 1 (actor *oid* in Fig. 3) maps the physical addresses of global variables to fixed virtual addresses (the reason for this is clarified in the next section. We describe how to instrument code and how identify global variables as well). The *data race detector*, *information flow security verifier*, and *backtracking points detector* components act upon these virtual addresses.

For the sake of brevity, the methods of stateless model checking of LTL formulae in DSCMC [48] and verifying information flow security [49] are not discussed in this paper.

## 4. Implementation and code instrumentation

Given a multi-threaded program, first the program must be instrumented with some code to communicate with DSCMC. The program communicates with DSCMC via functions defined in the *Pi* library. This section describes some implementation issues and code instrumentation.

When a stateless model checker (such as DSCMC, Inspect, etc.) re-executes the program under test, the runtime environment may change across re-executions. For instance, the operating system may not allocate the same identities to the threads. Moreover, in different runs, shared objects dynamically created (e.g., *mallocs*) may not reside in the same physical memory addresses. It is essential to handle these practical issues in stateless model checking [20]. We handle these issues in DSCMC the same as

Inspect [20].

According to [20], given the same external input, even across different runs, multiple threads in a program are always created in the same order. Using this fact, we can identify threads across two different runs by examining the order of thread creations. Each thread should register itself to the scheduler. If threads are created in the same sequential order in different re-execution, they will be registered in the same order. So we can easily assign the same identities for the same threads across different runs.

The shared objects are identified in the same way (from [20]): if two runs of the program have the same visible operation sequence, the shared objects will also be created with *malloc*, etc., in the same sequence. As a result, shared objects across different runs can be identified even though the actual objects may reside in different memory addresses at different runs. Actor *oid*, shown in Fig. 3, is responsible for registering shared objects and assigning virtual addresses to them.

DSCMC as well as other stateless model checkers [8, 24] eliminates inherent non-determinism in multithreaded execution by systematically exploring. When a bug is revealed, DSCMC exactly reproduces the execution that led to the bug.

The process of code instrumentation in DSCMC is fairly similar to Inspect. Appendix A shows how the program transforms at the instrumentation phase. The following are performed in code instrumentation (inspired from [8]):

- Function calls to the thread library routines are replaced with the function calls shown in Appendix A.
- Some code is added before thread starting/exiting points to notify the scheduler.
- For global objects, object registration code is added at the beginning of *main* function.
- After any operation that creates a new shared object, code is needed to register that object.
- DSCMC intercepts each read/write access on shared data objects by adding a wrapper around it.

To achieve the last step, we need to detect visible operations when a data object is updated. Doing this exactly is not decidable, as it amounts to context sensitive language reachability [8]. Inspect conservatively over-approximates this step by inter-procedural alias analyzing [27, 28]. Therefore, the important point is that the instrumentation must be safe for intercepting all the visible operations in the execution [8]. As the code is manually instrumented in the first prototype of DSCMC, we exploit the capabilities of compilers GCC [29] and LLVM [30] for alias analysis.

DSCMC uses a variety of techniques to address the state (and path) explosion problem. The DSCMC scheduler is non-preemptive so it allows a program to atomically execute large bodies of code. This search strategy is called preemption bounding [31], the intuition behind which is that in multi-threaded programs, many bugs are revealed by a few preemptions occurring in particular points in program execution [24]. The experimental results in [24, 31] showed the power of this approach. Besides, DSCMC uses dynamic partial order reduction [18, 25] and fair scheduling [1] techniques.

As we mentioned before, DSCMC expects programs to terminate under all fair schedules though these programs may not terminate under unfair schedules. In other words, non-termination under a fair schedule is potentially an error. Such programs are called fair-terminating [1]. The notion of fair-terminating programs has been inspired by applying the test harnesses in real-world concurrent programs. In practice, concurrent programs are tested with a suitable test harness that makes them fair-terminating. A fair scheduler eventually gives a chance of progressing to every thread in the program ensuring that the program makes progress towards the end of the test. Such a test harness can be created even for systems designed to "run forever" [1]. The DSCMC scheduling algorithm is accurately the same as [1]. For a more detailed discussion, please see [1].

The other source of non-determinism arises from calls to *random*() and *gettimeofday*(), which can return different values at different iterations of stateless model checking [24]. We expect the tester to avoid making such calls in the program under test. However, such calls might still be present in the code over which the tester has no control. We replace calls to such functions with predefined constant number at the code instrumentation phase.

At the instrumentation level, we partition the code to reduce the number of possible interleavings. Partitioning is based on the notion of dependent transitions. Informally, two transitions are independent if their order of execution is irrelevant. That is, either order results in the same program state, and none of the transitions can enable/disable the other one. As an example of dependent transitions, consider two threads updating the same shared variable, or entering the same critical section. Two threads updating their local variables, or reading the same shared variable are both examples of independent transitions [32].

Partitioning considers consecutive invisible operations with only one visible operation as a single operation. i.e., each partition of the code starts with a visible operation, and ends just before the next visible operation. Fig. 4 shows the code partitioning. Threads are scheduled upon their partitions. When a thread is scheduled, if it can progress, it continues executing until the end of its current partition. After reaching the end of a partition, it yields the processor to *Pi*, and *Pi* sends a request to the scheduler to schedule an enabled thread. If the thread holding the processor cannot progress, it prepares a message for the scheduler, which says the scheduler should choose another thread, and then sends this message via *Pi*.

In practice, a stateless model checker cannot identify or generate an infinite execution. Therefore, the user should set a large bound on the execution depth. This bound can be orders of magnitude greater than the maximum number of steps after which it is expected that the program will end. The model checker stops exploring the execution path (an iteration of the test) exceeding the bound, and reports a warning to the user [1]. Partitioning idea helps users to estimate the bound. For a program composed of *n* threads each of which is divided to $p_i$ partitions (after partitioning), the number of steps after that the program is expected to end is $\sum_{i=1}^{n} p_i$.

For the first prototype of DSCMC, we use the *erl_interface.h* library [34] to implement *Pi*. Currently, the *Pi* library is prepared for instrumenting code written in C and the POSIX *pthread* library [35]. Implementing the *Pi* is similar to integrating other code in different languages with an Erlang program [16, 17, 34].

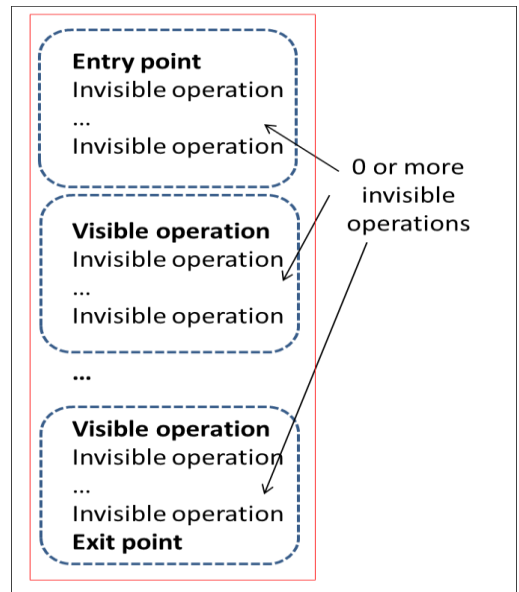

**Fig. 4.** Code partitioning

Consider the POSIX standard for threads [35]; function calls to the *pthread* library can be categorized in two types: (1) calls to functions that may cause a thread to yield the processor, such as to lock a mutex, wait on a condition variable or semaphore, etc. (2) function calls whereby a thread continues executing without waiting for a special source, such as to unlock a mutex, signal a semaphore, etc. We call the former case *waiting operations* and the latter case *non-blocking operations*. These two types are instrumented in a different manner.

Every program starts from an entry point, such as function *main* in C. The thread that contains the entry point is considered as the main thread and will have the first identity in the scheduler (i.e., *tid = 0*). When the program is started, the main thread registers itself in the scheduler via the message of the form *{n, Tid, Op, Oid}* and here, for the main function, *Tid = 0*, *Op = 'dc'*, and *Oid = 'dc'*. This message says to the scheduler, "I am the thread with identity *Tid*, and going to execute a non-blocking operation". In the above message, atom[2] *'n'* denotes the thread is going to execute a non-blocking operation and atom *'dc'* means *don't care*. However, generally, the message of the form *{Token, Tid, Op, Oid}* means the thread with identity *Tid* is going to execute operation *Op* on the global variable with identity *Oid* where *Token* can be either *'n'* or *'y'* for non-blocking and waiting operations respectively. On the other hand, all operations on the global variables are either

---

[2] Atom is a data type in Erlang. Please see [16].

*write* or *read* actions so variable *Op* is either *'r'* or *'w'* for *read* and *write* operations respectively. Such messages are sent to *Ti* by using functions of the *Pi* library as we described before. After receiving such a message, the scheduler chooses an enabled thread. If the selected thread is going to execute a waiting operation, it should send message *{pass}* to the scheduler after progressing. If the selected thread cannot progress it should yield the processor to *Pi* and send message *{yield}* to the scheduler. In the former scenario, the scheduler advances the program state and in the latter one, it should schedule another enabled thread. If a thread with a non-blocking operation is selected, the program need not send any messages to the scheduler and the scheduler advances the program state after scheduling. Fig. 5 shows the code instrumentation procedure in DSCMC. At the first step, it is necessary to identify shared variables so alias analysis is needed. Thereafter, all accesses (read/write) to these variables should be marked as visible operations. After that, we can partition the code so that each thread can yield the processor to *Pi* at the end of each partition. Here, we avoid describing implementation issues in too much detail however you can understand it by information existing in Appendix A and B. After partitioning, the code must be instrumented so as to be able to apply the DPOR algorithm. For this reason, each message sent to the scheduler contains information about the type of the thread action and the variable on which the thread is going to act. That is, actor *dpor* using the variables *Op* and *Oid* in message *{Token, Tid, Op, Oid}* can detect independent transitions. If the current state contains two or more enabled transitions, the scheduler sends their *Ops* and *Oids* as messages of the form *{Op, Oid}* to *dopr*, and then *dpor* eliminates the independent transitions. After that, if the state still contains at least two enabled transitions, that state is considered as a backtracking point and its information is captured by *btset*.

At the fifth step, the identities must be allocated to the threads and global variables. In this regard, at each point of the code where a new thread is created, a new identity is assigned to it. Whenever the thread is going to send a message for the scheduler, it should use this identity to prepare the message in the manner described above (i.e., as variable *Tid* in the message tuple).

DSCMC can be used in two configurations: (1) verification, and (2) security verification. In the first configuration, DSCMC can detect deadlocks, livelocks, data races, and as well as check the LTL formulae. The second configuration is dedicated to verify information flow security. Code instrumentation is differently performed for each configuration. The reminder of instrumenting process (i.e., after the fifth step) is shown with the number *1* for the first configuration and *2* for the second one in Fig. 5.

The first configuration should detect data races, which is completely described in the next subsection. Thereafter, for checking LTL formulae (if there exist) code must be instrumented. Checking LTL formulae is not discussed in this paper, for precise information, please see [48]. The second configuration is also elaborated in [49] and it is not addressed in this paper.

## 4.1 Detecting data races

Data race detection is essential for debugging multithreaded programs and assuring their correctness. The data race detection problem is computationally hard so that there is no single universal technique for efficiently performing the task [36]. DSCMC is able to detect data races due to systematic testing of programs by method proposed in this section.

When the following conditions hold, a data race occurs: (1) at least two threads of a single process access the same memory location concurrently, (2) at least one of the accesses is for writing, and (3) the threads are not using any exclusive locks that controls accesses to that memory. At the instrumentation phase for the first configuration (i.e., the sixth step), we must identify shared variables that are not controlled with any exclusive locks (we denote these variables with *RV*). Thereafter, the code is instrumented such that the program behaves as follows:

Whenever a thread is going to access to any variable in *RV*, a message of the form *{cnode, drace, {w, Oid}}* (for write access) or *{cnode, drace, {r, Oid}}* (for read access) is sent to *Ti* by actor *Pi* where variable *Oid* is identity of that *RV* variable. *Ti* according to the message header (i.e., *cnode, drace*) finds that the received message should deliver to the actor detecting data races. *Ti* removes the header, and sends the message of the form *{w, Oid}* or *{r, Oid}* to the race detector. The behavioral function of this actor is shown in Fig. 6.

DSCMC's *data race detector*, shown in Fig 1, is a collection of actors running in concurrent where the *master* actor is the actor with behavioral function *race_detector* shown in Fig. 6. This actor creates a small and light actor for each *RV*, which behaves corresponding to function *detect* in Fig. 6 (that are called *workers*). During stateless model checking, when an *RV* is initialized, a message of the form *{dc, reg_obj_id, Adrs}* is sent to actor *oid* where *Adrs* is the physical address of the variable. The *oid* actor assigns an identity to this variable and sends message *{reg_obj_id, Oid}* to the master where *Oid* is the corresponding identity of *Adrs*. After that, whenever *oid* receives a message pertaining to the race detector component, it maps the address to the corresponding *Oid*, and sends that message to the master. During this operation, *oid* performs dynamic alias analysis as well. The analysis is precise and based on physical addresses.

As mentioned above, when the master receives a message about registering a new variable, it creates a worker to monitor accesses to that variable (Fig. 6, line 8). At runtime, when a thread is going to write on an *RV*, the master receives a message of the form shown at line 3, and then forwards it to the worker monitoring that variable. Then, the worker increments the number of writer threads at line 13 (i.e., the number of the threads that are going to write on that variable). When any of these threads ends its writing operation, the master receives a relating message to this event at line 5, and forwards it to the related worker. Thereafter, the worker decrements the number of writer threads. In the same way, the worker increments and decrements the number of reader threads (lines 4-6). Workers use function *check* (line 19) to detect data races. The function checks the existence of a race situation at line 21 such that if both the number of reader and the number of writer threads on an *RV* are greater than zero, a data race situation is revealed. After detecting a race situation, the path where the data race has occurred is saved for further inspection. The user can precisely re-execute the path using the *re-executing component* (Fig. 1). When an iteration of stateless model checking ends, the master actor receives message *finish* from the controller at line 9 whereby destroys all the workers.

Using the proposed method, DSCMC reports a race if and only if a racing scenario occurs. A question that may come into mind is "the method may cause to create many Erlang processes, is it efficient?" let us refer you to Mr. Armstrong's book [16] where he has shown Erlang programs can be made from thousands to millions of extremely lightweight processes that can even run on a single processor. His experiment on the computer with 2.40GHz Intel Celeron and 512MB of memory running Ubuntu Linux has shown that spawning 20,000 processes took an average of 3.5 μs/process of CPU time. You can easily repeat this experiment. The fact is that Erlang processes are indeed light.

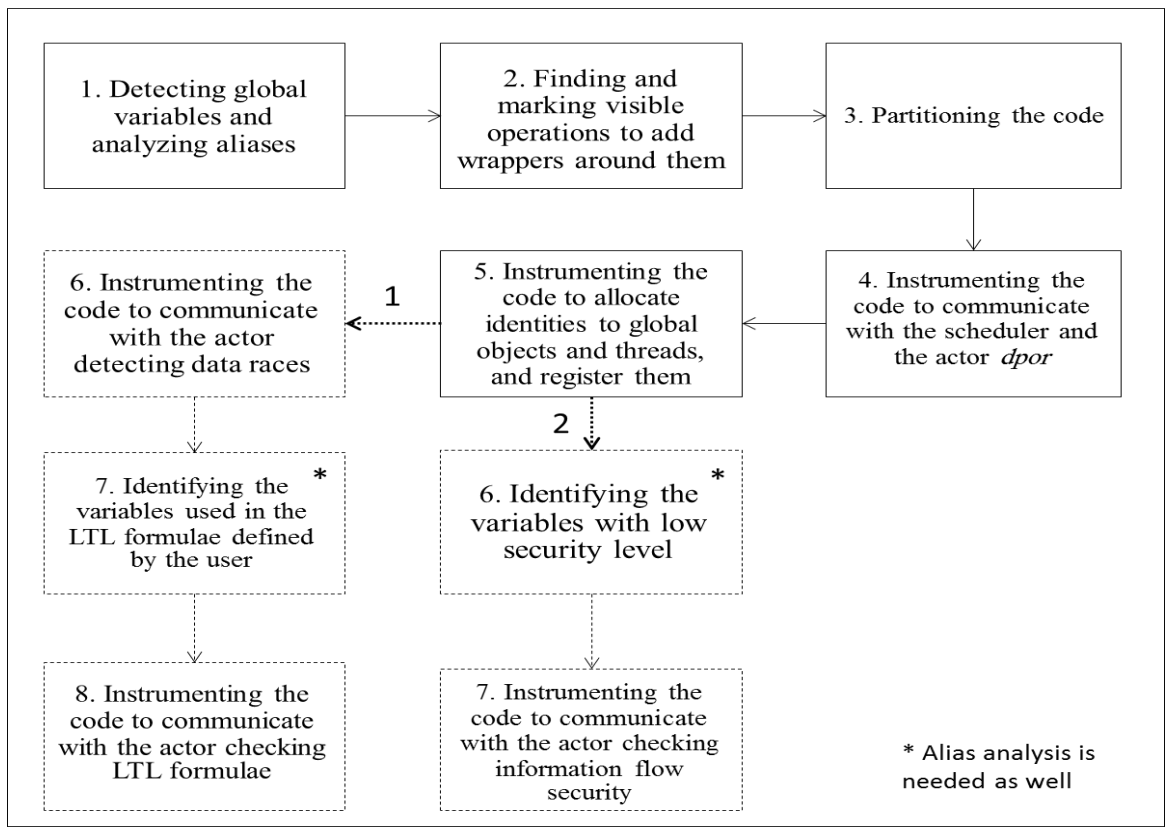

**Fig. 5.** The code instrumentation procedure in DSCMC

```
1. race_detector(Names) ->
2.   receive
3.           {w, S} -> list_to_atom(S) ! w , race_detector(Names);
4.           {r, S} -> list_to_atom(S) ! r , race_detector(Names);
5.        {wrote, S} -> list_to_atom(S) ! e_w , race_detector(Names);
6.       {readed, S} -> list_to_atom(S) ! e_r , race_detector(Names);
7.    {reg_obj_id, S} -> N = list_to_atom(S),
8.                  register (N, spawn(fun()-> detect(0, 0) end)),  race_detector( [N|Names] );
9.           finish -> spawn(fun()-> lists:foreach(fun(X)->  exit(whereis(X), kill) end, Names ) end)
10.  end.
11. detect (R, W) ->
12.   receive
13.     w -> check (R, W+1), detect (R, W+1);
14.     r -> check (R+1, W), detect (R+1, W);
15.    e_r -> detect (R-1, W);
16.    e_w -> detect (R, W-1)
17.  end.
18.
19. check (R, W) ->
20.   if
21.     (W > 0 andalso R > 0) ->  % data race situation… saving the path
22.                true  -> true
23.   end.
```

**Fig. 6.** The behavioral function of actors detecting data races

## 5. Evaluation

We have used the Spin model checker [37] and the Rebeca modeling language [11, 12] to verify DSCMC. As DSCMC has an actor-based architecture and has been implemented using an actor-based programming language, we have decided to choose the Rebeca language to model it. We should convert Rebeca models into Promela [38] so as to use Spin. Spin is a well-known and powerful model checker to verify LTL properties of the models written in Promela. We have used the R2P tool, proposed in [39] for model converting.

DSCMC has been tested via several examples, some of which are included in Appendix B. As code is manually instrumented at the present time, we have not been able to test DSCMC in large programs yet. Nevertheless prior stateless model checkers have shown their capability to verify large realistic programs [23, 26, 33, 40], hence DSCMC is expected to inherit this capability due to its stateless nature. Moreover, DSCMC has a parallel architecture because of which it should achieve better results while verifying large programs.

Note that when DSCMC runs on a multi-core processor, a better performance should be gained; because of the parallel architecture, the overhead of runtime monitoring imposed by the data race detectors, actors checking LTL formulae, the *dpor* component, etc. can be distributed on different cores with minimal side effect (because of the independence of these actors). Fortunately, the concurrency model of Erlang—separate processes (actors) communicating via message passing without any shared memory —naturally transfers to multicore processors in a way that is completely transparent to the programmer, so that we can run DSCMC as well as other Erlang programs on more powerful hardware without having to redesign [17]. In addition, distributed nature of the tool causes the process of the state space exploration to be distributed among several computational nodes so the time of the process should dramatically reduce while DSCMC is running in a distributed environment.

## 6. Related work

As far as we know, there have been three *stateless* model checkers, namely VeriSoft [5, 40], CHESS [24, 33], and Inspect [20, 26] prior to DSCMC [41]. VeriSoft is optimized for analyzing *multi-process* applications. It can analyze systems composed of processes with hundreds of thousands of lines of code. However, Inspect and CHESS are concerned with verifying multi-threaded programs. Therefore, we compare DSCMC with these two tools. CHESS [33] was developed in Microsoft Research in 2007 [24]. Inspect [26] was developed at the University of Utah [42] in 2008. Several researches were performed to improve stateless model checking by their developers, including [1, 8, 19, 20, 23, 31].

On the downside, Inspect and VeriSoft can only handle programs that can terminate in a finite number of steps. Such programs terminate under all executions, and equivalently have acyclic state spaces though most realistic concurrent programs have cyclic state spaces [1]. A technique of fair stateless model checking has been introduced in [1] for effectively exploring the state spaces of nonterminating programs, which has been implemented in CHESS. Without fair searching technique [1], non-termination and cyclic state spaces present a significant obstacle to stateless model checkers (such as Inspect and VeriSoft) so these tools are limited to the verification of safety properties on terminating programs (such as deadlocks and assertions violations).

In comparing CHESS with Inspect, the strength of CHESS is that it is able to detect and prune unfair cycles in the program state space so it can be used to detect the problems pertaining to fair cycles (e.g., livelocks). The strengths of Inspect are that it is distributed [23, 43], and supports POSIX threads. But, unfortunately, none of the tools support verifying LTL formulae. We have proposed a novel method for stateless model checking of LTL formulae in DSCMC [48]. Besides, DSCMC have been developed so that it can check information flow security in concurrent programs. Table 1 compares capabilities of DSCMC with Inspect and CHESS.

**Table 1.** Comparing DSCMC with Inspect and CHESS

| Features | DSCMC | Inspect | CHESS |
|---|---|---|---|
| Checking LTL formulae | Yes | No | No |
| Detecting deadlocks | Yes | Yes | Yes |
| Detecting livelocks | Yes | No | Yes |
| Detecting data races | Yes | Yes | Yes |
| Detecting assertions violations | Yes | Yes | Yes |
| Detecting and pruning unfair cycles | Yes | No | Yes |
| Checking information flow security | Yes | No | No |
| Distribution | Yes | Yes | No |
| Parallelism | Yes | No | No |
| Supporting different programming languages | Yes | No | No |

# 7. Conclusions

This paper presents a new stateless code model checker with an actor-based nature, called DSCMC. DSCMC is used for systematic testing concurrent programs under specific input. The required tasks of stateless model checking are dispatched among DSCMC's actors running in parallel. DSCMC executes programs in their realistic runtime environment under a special scheduler.

Due to the actor-based architecture, DSCMC proposes a new method for stateless checking of LTL formulae [48]. Besides, data race detection, checking information flow security, etc. are based on the Actor model. DSCMC is suited to verify concurrent programs. It can reveal common concurrency errors (such as deadlocks, livelocks, and data races) in concurrent programs without any false alarms.

To use DSCMC, first, program code must be instrumented. Currently, code is manually instrumented in DSCMC. Therefore, in the future, code instrumentation must be automated so that DSCMC can be utilized in verifying large programs. It is noteworthy that the automatic code instrumentation had better perform on the intermediate code of the common compilers, such as GIMPLE [29], and LLVM IR [30]. By employing these representations, we would exploit the strong features of these compilers to analyze code at the instrumentation phase. Moreover, at the intermediate level, all the high-level control flow structures, including exception handling, and loops, nested functions, and breaks expressions are linearized down into a three-address form. Working with intermediate representation allows our analysis to be focused more on control flow information and data modifications rather than spend effort to analyze complex constructs of the programming languages [44].

Existing stateless model checkers cannot cope with the non-determinism from user input. They assume that there is specific input for the concurrent program under test, and explores the non-determinism due to concurrency. To remedy this problem, the method can be improved by combining test generation techniques (such as white-box fuzz testing [45] and symbolic execution [46]) with DSCMC's components to cover more execution paths. The method proposed by Godefroid *et al.* in [45, 47] can be effective for this purpose.

## Acknowledgement

We are grateful to Iran National Science Foundation (INSF) for financial support of this research.

# Appendix A . Code transformation

This appendix shows how function calls to the *pthread* library are transformed at the instrumentation level. At this phase, we transform *waiting* function calls to *trying* function calls, as can be seen in the following table.

| | Function calls to pthread- Before instrumentation | Function calls to pthread- After instrumentation |
|---|---|---|
| 1 | pthread_create (&thread_name, Null, thread_function, Null); | pthread_mutex_lock (&id_mutex);<br>params_t[id].tid = id;   *// assign an identity to new thread*<br>send_create();       *//send message {create} to the scheduler*<br>*/* pass the identity as the argument to new thread */*<br>pthread_create (&thread_name, Null, thread_function, ¶ms_t[id])<br>id = id + 1;<br>pthread_mutex_unlock (&id_mutex);<br>send_stat("n",myTid); *//send message {n, myTid, dc, dc} to the scheduler*<br>sem_post (&rcv_sem);    *// yield the processor to Pi*<br>sem_wait (&permit_to_exe[myTid]); *// wait for taking the processor* |
| 2 | Pthread_join (thread_name, Null); | ch:<br>sem_wait (&permit_to_exe[myTid]); *// wait for taking the processor*<br>if (pthread_tryjoin_np(thread_name, NULL) == EBUSY){<br>    sem_post(&rcv_sem);  *// yield the processor to Pi*<br>    send_yield();           *// send message {yield} to the scheduler*<br>    goto ch;<br>}<br>send_pass();                       *// send message {pass} to the scheduler* |
| 3 | Pthread_mutex_lock (&pthread_mutex_t_name); | ch:<br>sem_wait (&permit_to_exe[myTid]);<br>if ( pthread_mutex_trylock(&pthread_mutex_t_name)== EBUSY){<br>    sem_post (&rcv_sem);     send_yield();        goto ch;<br>  }     send_pass(); |
| 4 | sem_wait (&sem_t_name); | ch:<br>sem_wait (&permit_to_exe[myTid]);<br>sem_getvalue (&sem_t_name, &val);<br>if (val == 0){<br>    sem_post (&rcv_sem);     send_yield();       goto ch;<br>}<br>sem_wait (&sem_t_name);      send_pass(); |
| 5 | pthread_cond_wait(&pthread_cond_t_name, &pthread_mutex_t_name);<br><br>All condition variables are converted to the integer variable.<br>As can be seen in the transformed code, condition variable *pthread_cond_t_name* is replaced with integer variable *conditionVar*. Another integer variable used to emulate the condition variable is *conditionVarW*.<br>Note that the DSCMC scheduler is non-preemptive so there is no concern about interruption while a thread is executing the following code on the front column.<br>We use these integer variables to emulate *pthread_cond_signal* as well (i.e., in the next row). | if (conditionVarsW > 0 && conditionVar == 1){   *// these are integer variables*<br>    pthread_mutex_unlock(&pthread_mutex_t_name);<br>}<br>else{<br>    pthread_mutex_unlock (&pthread_mutex_t_name);<br>    send_no_dead(myTid*);*   *//sending message {no_dead} to the scheduler*<br>   conditionVarsW += 1;<br>  ch_1:<br>  if(conditionVar == 0){<br>    send_yield(); sem_post (&rcv_sem);<br>    sem_wait (&permit_to_exe[myTid]);    goto ch_1;<br>  }<br>  ch_2:<br>  if (pthread_mutex_trylock(&pthread_mutex_t_name)== EBUSY){<br>    sem_post (&rcv_sem); send_yield();<br>    sem_wait (&permit_to_exe[myTid]);       goto ch_2;<br>  }<br>  conditionVarsW -= 1;    conditionVar = 0;<br>}<br>  send_pass(); |
| 6 | pthread_cond_signal   (&pthread_cond_t_name); | if(conditionVarsW > 0)  conditionVars = 1; |

# Appendix B.    Examples

This appendix gives some examples of verifying multi-threaded programs by DSCMC.

## B.1. Detecting data races

The data race problem is well-known vulnerability of concurrent programs, which can be the origin of some attacks. This section presents a multi-threaded program with the data race problem. The program is verified by DSCMC whereby the data race situation is detected. In this example, the race situation is caused by unprotected thread accesses on the shared variable *flag*. Fig. 7 is the code example written in C and *pthread*. The instrumented code for this program is shown in Fig. 8, where bolded code parts were added during instrumenting. Fig. 9 shows a scenario that has caused the data race situation, detected by DSCMC. The output contains the order of thread scheduling, which is a trace file used by the *re-executing component*. This trace file shows that if thread 0 (i.e., the main thread: thread which contains function *main*) is scheduled for four times, a data race will occur. As can be seen in Fig. 8, the thread 0, in its second partition[3], creates thread 1 (i.e., the thread with function *thread1*), next creates thread 2 (i.e., the thread with function *thread2*) at its own third partition. When thread 1 is created, it wants to read from *flag*. But thread 0 is scheduled, and creates thread 2. When thread 2 is created, it wants to write on *flag*. At present, thread 1 is going to read from *flag* and thread 2 is going to write on so a data race has occurred.

```
#include <pthread.h>
int flag;
void * thread1(void * arg)
{
      if  (flag == 1)    printf ("flag is set\n" );
       return 0;
}
void * thread2(void * arg)
{
      flag = 0;
}

int main()
{ pthread_t t1, t2;
  flag = 0;  //initialization
  pthread_create(&t1, 0, thread1, 0 );
  pthread_create(&t2, 0, thread2, 0);
  pthread_join(t1, 0);
  pthread_join(t2, 0);
  return 0;
}
```

**Fig. 7.** A multi-threaded program with the data race problem

---

[3] The notion of the *partition* is elaborated in the paper (see Fig. 4).

```c
#include <pthread.h>
#include <stdio.h>
#include <malloc.h>
#include <errno.h>
#include "nodec.h"
int flag;
void * thread1(void * arg)
{
  pthread_mutex_lock (&id_mutex);
  struct pthread_params* temp =
          (struct pthread_params*)arg;
  int myTid;
  myTid = temp->tid;
  pthread_mutex_unlock (&id_mutex);
  send_stat_exe("n", myTid, &flag, "r");
  read_shobj(&flag);
  sem_post (&rcv_sem);
  sem_wait (&permit_to_exe[myTid]);

  if (flag == 1) {
       readed (&flag);
       printf ("flag is set\n");
  }
  send_end(myTid);
  sem_post (&rcv_sem);
  return 0;
}

void * thread2(void * arg)
{
  pthread_mutex_lock (&id_mutex);
  struct pthread_params* temp =
          (struct pthread_params*)arg;
  int myTid;
  myTid = temp->tid;
  pthread_mutex_unlock (&id_mutex);
  send_stat_exe("n", myTid, &flag, "w");
  write_shobj(&flag);
  sem_post (&rcv_sem);
  sem_wait (&permit_to_exe[myTid]);

  flag = 0;
  wrote(&flag);
  send_end(myTid);
  sem_post (&rcv_sem);
  return 0;
}
int main(int argc, char *argv[])
{
  if (!argc)
  {
    printf("Missing input argument!\n");
    printf("Example: e1@10.0.2.15 \n");
    return 1;
  }
num_of_threads = 3;
maininit(argv[1]);
struct pthread_params params_t [num_of_threads];
int myTid = id;
id = id +1;
send_stat("n",myTid);
sem_post (&rcv_sem);
sem_wait (&permit_to_exe[myTid]);

pthread_t t1, t2;
flag = 0;
reg_oid (&flag);
pthread_mutex_lock (&id_mutex);
params_t[id].tid = id;
send_create();
pthread_create(&t1, NULL,  thread1, ¶ms_t[id] );
 id = id + 1;
pthread_mutex_unlock (&id_mutex);
send_stat("n",myTid);
sem_post (&rcv_sem);
sem_wait (&permit_to_exe[myTid]);
pthread_mutex_lock (&id_mutex);
params_t[id].tid = id;
send_create();
pthread_create(&t2, NULL,  thread2, ¶ms_t[id] );
 id = id + 1;
pthread_mutex_unlock (&id_mutex);
send_stat("y",myTid);
sem_post (&rcv_sem);
ch1:
sem_wait (&permit_to_exe[myTid]);
if (pthread_tryjoin_np(t1, NULL) == EBUSY){
     sem_post(&rcv_sem); send_yield(); goto ch1;
}
send_pass();
send_stat("y",myTid);
sem_post (&rcv_sem);
ch2:
sem_wait (&permit_to_exe[myTid]);
if (pthread_tryjoin_np(t2, NULL) == EBUSY){
     sem_post(&rcv_sem); send_yield(); goto ch2;
}
send_pass();
destroy_erl();
send_end(myTid);
return 0;   }
```

**Fig. 8.** The instrumented code of the program in Fig. 7

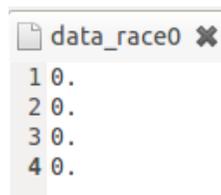

**Fig. 9.** The trace file of the path that has caused the data race.

### B.2. Detecting deadlocks

Deadlock is a common problem in thread programming. DSCMC can precisely detect deadlocks. A program with deadlock problem is shown in Fig. 10, written in C and *pthread*. Fig. 11 shows the instrumented code. In Fig. 12, DSCMC detects the existing deadlock scenario in the program. Fig. 12 is the trace file of the path that caused the deadlock. As DSCMC uses the *dynamic partial order reduction* algorithm, another deadlock scenario that begins with thread 2 at the third scheduling point is omitted. The trace file in Fig. 12 shows that when thread 0 is scheduled two times, then thread 1 is scheduled and executes its first partition, next thread 2 is scheduled and executes its own first partition too, a deadlock problem occurs.

```c
#include <pthread.h>
pthread_mutex_t  a, b;
int counter;
void * thread1(void * arg)
{
    pthread_mutex_lock(&a);
    pthread_mutex_lock(&b);
    counter++;
    pthread_mutex_unlock(&b);
    pthread_mutex_unlock(&a);
    return 0;
}
void * thread2(void * arg)
{   pthread_mutex_lock(&b);
    pthread_mutex_lock(&a);
    counter--;
    pthread_mutex_unlock(&a);
    pthread_mutex_unlock(&b);
}
int main()
{
    pthread_t t1, t2;       counter = 0;
    pthread_mutex_init(&a, 0);
    pthread_mutex_init(&b, 0);
    pthread_create(&t1, 0, thread1, 0);
    pthread_create(&t2, 0, thread2, 0);
    pthread_join(t1, 0);
    pthread_join(t2, 0);
    return 0;
}
```

**Fig. 10.** A multi-threaded program with deadlock

```c
#include <pthread.h>
#include <stdio.h>
#include <malloc.h>
#include <errno.h>
#include "nodec.h"
pthread_mutex_t  a, b;
int counter;
void * thread1(void * arg)
{
  pthread_mutex_lock (&id_mutex);
  struct pthread_params* temp =
            (struct pthread_params*)arg;
  int myTid;
  myTid = temp->tid;
  pthread_mutex_unlock (&id_mutex);
  send_stat_exe("y", myTid, &a, "w");
  sem_post (&rcv_sem);
ch1:
  sem_wait (&permit_to_exe[myTid]);
  if ( pthread_mutex_trylock(&a)== EBUSY){
     sem_post (&rcv_sem); send_yield(); goto ch1;
  }
  send_pass();
  send_stat_exe("y", myTid, &b, "w");
  sem_post (&rcv_sem);
ch2:
  sem_wait (&permit_to_exe[myTid]);
  if ( pthread_mutex_trylock(&b)== EBUSY){
     sem_post (&rcv_sem); send_yield(); goto ch2;
  }
  send_pass();
  send_stat_exe("n", myTid, &counter, "w");
  sem_post (&rcv_sem);
  sem_wait (&permit_to_exe[myTid]);
  counter = counter + 1;
  send_stat_exe("n", myTid, &b, "w");
  sem_post (&rcv_sem);
  sem_wait (&permit_to_exe[myTid]);
  pthread_mutex_unlock(&b);

  send_stat_exe("n", myTid, &a, "w");
  sem_post (&rcv_sem);
  sem_wait (&permit_to_exe[myTid]);
  pthread_mutex_unlock(&a);

  send_end(myTid);
  sem_post (&rcv_sem);
  return 0;   }

void * thread2(void * arg)
{
  pthread_mutex_lock (&id_mutex);
  struct pthread_params* temp =
            (struct pthread_params*)arg;
  int myTid;
  myTid = temp->tid;
  pthread_mutex_unlock (&id_mutex);
  send_stat_exe("y", myTid, &b, "w");
  sem_post (&rcv_sem);
```

**Fig. 11.** The instrumented code of the program shown in Fig. 10

```
ch1:
 sem_wait (&permit_to_exe[myTid]);
 if ( pthread_mutex_trylock(&b)== EBUSY){
     sem_post (&rcv_sem); send_yield(); goto ch1;
 }

 send_pass();

 send_stat_exe("y", myTid, &a, "w");
 sem_post (&rcv_sem);
 ch2:
 sem_wait (&permit_to_exe[myTid]);

 if ( pthread_mutex_trylock(&a)== EBUSY){
     sem_post (&rcv_sem); send_yield(); goto ch2;
 }

 send_pass();
 send_stat_exe("n", myTid, &counter, "w");
 sem_post (&rcv_sem);
 sem_wait (&permit_to_exe[myTid]);

 counter = counter - 1;
 send_stat_exe("n", myTid, &a, "w");
 sem_post (&rcv_sem);
 sem_wait (&permit_to_exe[myTid]);
 pthread_mutex_unlock(&a);

 send_stat_exe("n", myTid, &b, "w");
 sem_post (&rcv_sem);
 sem_wait (&permit_to_exe[myTid]);
 pthread_mutex_unlock(&b);
 send_end(myTid);
 sem_post (&rcv_sem);
 return 0;
}
int main(int argc, char *argv[])
{
 if (!argc)
 {
   printf("Missing input argument!\n");
   printf("Example: e1@10.0.2.15 \n");
   return 1;
 }
 num_of_threads = 3;
 maininit(argv[1]);
 struct pthread_params params_t [num_of_threads];
 int myTid = id;
 id = id +1;

 send_stat("n",myTid);
 sem_post (&rcv_sem);
 sem_wait (&permit_to_exe[myTid]);
 pthread_t t1, t2;
 counter = 0;
 reg_oid (&counter);
 pthread_mutex_init(&a, 0);
 reg_oid (&a);
 pthread_mutex_init(&b, 0);
 reg_oid (&b);
 pthread_mutex_lock (&id_mutex);
 params_t[id].tid = id;
 send_create();
 pthread_create(&t1, NULL, thread1, ¶ms_t[id] );
  id = id + 1;
 pthread_mutex_unlock (&id_mutex);
 send_stat("n",myTid);
 sem_post (&rcv_sem);
 sem_wait (&permit_to_exe[myTid]);
 pthread_mutex_lock (&id_mutex);
 params_t[id].tid = id;
 send_create();
 pthread_create(&t2, NULL, thread2, ¶ms_t[id] );
  id = id + 1;
 pthread_mutex_unlock (&id_mutex);
 send_stat("y",myTid);
 sem_post (&rcv_sem);
 ch1:
 sem_wait (&permit_to_exe[myTid]);
 if (pthread_tryjoin_np(t1, NULL) == EBUSY){
     sem_post(&rcv_sem); send_yield(); goto ch1;
 }
 send_pass();
 send_stat("y",myTid);
 sem_post (&rcv_sem);
 ch2:
 sem_wait (&permit_to_exe[myTid]);
 if (pthread_tryjoin_np(t2, NULL) == EBUSY){
     sem_post(&rcv_sem); send_yield(); goto ch2;
 }
 send_pass();
 destroy_erl();
 send_end(myTid);
 return 0;
}
```

**Fig. 11.** The instrumented code of the program shown in Fig. 10 - *continuation*

```
bt_1_deadlock ✖
1 0.
2 0.
3 1.
4 2.
```

**Fig. 12.** The trace file of the path that has caused the deadlock

### B.3. Detecting livelocks

Another common problem in multi-threaded programs is known as livelocks. To illustrate the problem, consider the program in Fig. 13 (from [1]). The program is a variation of the dining philosophers example with two threads *Phil1* and *Phil2* trying to acquire two resources *fork1* and *fork2*. *Phil1* acquires *fork1* and then attempts to acquire *fork2* without blocking. If this attempt fails, then it releases *fork1* and retries. *Phil2* tries to acquire the two resources in the reverse. The repeated execution of the transition sequence *Phil1: Acquire(fork1), Phil2: Acquire(fork2), Phil1: TryAcquire(fork2), Phil2: TryAcquire(fork1), Phil1: Release(fork1), Phil2: Release(fork2)* is a one of the possible livelock scenarios in the program. We implement the pseudo code shown in Fig. 13. The instrumented code of the implemented program is shown in Fig. 14. Fig. 15 shows some possible livelocks in the program detected by DSCMC.

```
Phil1:
while ( true ) {
    Acquire ( fork1 );
    if ( TryAcquire ( fork2 ) )
        break;
    Release ( fork1 );
}
// eat
Release ( fork1 );
Release ( fork2 );

Phil2:
while ( true ) {
    Acquire ( fork2 );
    if ( TryAcquire ( fork1 ) )
        break;
    Release ( fork2 );
}
// eat
Release ( fork2 );
Release ( fork1 );
```

**Fig.13.** Pseudo code of a program with livelocks

```c
#include <stdlib.h>
#include <pthread.h>
#include <stdio.h>
#include <errno.h>
#include <sys/types.h>
#include "nodec.h"

#define NUM_THREADS 2
pthread_mutex_t f[NUM_THREADS];
pthread_t tids[NUM_THREADS];
void * Philosopher(void * arg){
   pthread_mutex_lock (&id_mutex);
   struct pthread_params* temp =
           (struct pthread_params*)arg;
   int myTid;
   int i;
   i = (int) temp->user_param;
   myTid = temp->tid;
   pthread_mutex_unlock (&id_mutex);
   send_stat("n",myTid);    sem_post (&rcv_sem);
   sem_wait (&permit_to_exe[myTid]);
   while(1) {
     send_stat_exe("y",myTid,&f[i],"w");
     sem_post (&rcv_sem);
     chk_2:
     sem_wait (&permit_to_exe[myTid]);
     if ( pthread_mutex_trylock(&f[i])== EBUSY){
       sem_post (&rcv_sem); send_yield(); goto chk_2;
     }
     send_pass();
     printf("take F %d: %d\n\r",i, myTid);
   send_stat_exe("n",myTid, &f[(i+1)%NUM_THREADS], "w");
     sem_post (&rcv_sem);
     sem_wait (&permit_to_exe[myTid]);
     if (pthread_mutex_trylock(&f[(i+1)%
              NUM_THREADS])!= EBUSY) {
              printf("take F %d: %d\n\r",
                 (i+1)%NUM_THREADS, myTid);
             break;
     }
     send_stat_exe("n",myTid,&f[i],"w");
     sem_post (&rcv_sem);
     sem_wait (&permit_to_exe[myTid]);
     printf("release F %d: %d\n\r",i, myTid);
     pthread_mutex_unlock (&f[i]);
  }
  printf("I'm Ph %d & I'm eating now...\n",i);
  send_stat_exe("n",myTid,&f[i],"w");
  sem_post (&rcv_sem);
  sem_wait (&permit_to_exe[myTid]);
  printf("release F %d: %d\n\r",i, myTid);
  pthread_mutex_unlock (&f[i]);
  send_stat_exe("n",myTid,&f[(i+1)%NUM_THREADS],"w");
  sem_post (&rcv_sem);
  sem_wait (&permit_to_exe[myTid]);
  printf("release F %d: %d\n\r",
          (i+1)%NUM_THREADS, myTid);
  pthread_mutex_unlock (&f[(i+1)%NUM_THREADS]);
  send_end(myTid);
  sem_post (&rcv_sem);
}

int main(int argc, char *argv[]){
if (!argc)
{
printf("Missing input argument!\n");
printf("Example: e1@10.0.2.15 \n");
return 1;
}
  num_of_threads = NUM_THREADS+1;
  maininit(argv[1]);

int myTid = id;
id = id +1;
send_stat("n",myTid);
sem_post (&rcv_sem);
sem_wait (&permit_to_exe[myTid]);

int i;
  for (i = 0; i < NUM_THREADS; i++){
    pthread_mutex_init(&f[i], NULL);
    reg_oid(&f[i]);
}
  struct pthread_params params_t [num_of_threads];

  for (i = 0; i < NUM_THREADS; i++){
    pthread_mutex_lock (&id_mutex);
    params_t[id].user_param = (void*)i;
    params_t[id].tid = id;
    send_create();
   pthread_create(&tids[i], NULL,  Philosopher,
          ¶ms_t[id] );
    id = id + 1;
    pthread_mutex_unlock (&id_mutex);

if(i<NUM_THREADS-1){
send_stat("n",myTid);
sem_post (&rcv_sem);
sem_wait (&permit_to_exe[myTid]);
else send_stat("y",myTid);
 }
sem_post (&rcv_sem);
chk_3:
sem_wait (&permit_to_exe[myTid]);

 for (i = 0; i < NUM_THREADS; i++){
   if (pthread_tryjoin_np(tids[i], NULL) == EBUSY){
     sem_post(&rcv_sem); send_yield(); goto chk_3;
    }
 }
send_pass();
 for (i = 0; i < NUM_THREADS; i++){
   pthread_mutex_destroy(&f[i]);
 }
destroy_erl();
printf(" destroy............ \n");
send_end(myTid);
 return 0;
}
```

**Fig. 14.** The implemented and instrumented code of Fig. 13

| bt_3_livelock | bt_5_livelock | bt_6_livelock | bt_10_livelock | bt_12_livelock | bt_13_livelock | bt_15_livelock |
|---|---|---|---|---|---|---|
| 1 0. | 1 0. | 1 0. | 1 0. | 1 0. | 1 0. | 1 0. |
| 2 0. | 2 0. | 2 0. | 2 0. | 2 0. | 2 0. | 2 0. |
| 3 1. | 3 1. | 3 1. | 3 1. | 3 1. | 3 1. | 3 1. |
| 4 1. | 4 1. | 4 1. | 4 1. | 4 1. | 4 1. | 4 1. |
| 5 2. | 5 2. | 5 2. | 5 2. | 5 2. | 5 2. | 5 2. |
| 6 2. | 6 2. | 6 2. | 6 2. | 6 2. | 6 2. | 6 2. |
| 7 2. | 7 2. | 7 2. | 7 2. | 7 2. | 7 2. | 7 2. |
| 8 1. | 8 1. | 8 1. | 8 2. | 8 2. | 8 2. | 8 2. |
| 9 1. | 9 1. | 9 1. | 9 2. | 9 2. | 9 2. | 9 2. |
| 10 2. | 10 2. | 10 2. | 10 1. | 10 1. | 10 1. | 10 1. |
| 11 1. | 11 1. | 11 1. | 11 1. | 11 1. | 11 1. | 11 2. |
| 12 2. | 12 2. | 12 2. | 12 1. | 12 1. | 12 1. | 12 2. |
| 13 2. | 13 2. | 13 2. | 13 2. | 13 2. | 13 2. | 13 1. |
| 14 1. | 14 2. | 14 2. | 14 1. | 14 2. | 14 2. | 14 1. |
| 15 1. | 15 2. | 15 2. | 15 1. | 15 2. | 15 2. | 15 2. |
| 16 2. | 16 1. | 16 1. | 16 2. | 16 1. | 16 1. | 16 1. |
| 17 1. | 17 1. | 17 2. | 17 1. | 17 1. | 17 2. | 17 1. |
| 18 2. | 18 1. | 18 2. | 18 2. | 18 1. | 18 2. | 18 1. |
| 19 2. | 19 2. | 19 1. | 19 2. | 19 2. | 19 1. | 19 2. |
| 20 1. | 20 1. | 20 1. | 20 1. | 20 1. | 20 1. | 20 2. |
| 21 1. | 21 1. | 21 2. | 21 1. | 21 1. | 21 2. | 21 2. |
| 22 2. | 22 2. | 22 1. | 22 2. | 22 2. | 22 1. | 22 1. |
| 23 1. | 23 1. | 23 1. | 23 1. | 23 1. | 23 1. | 23 1. |
| 24 1. | 24 1. | 24 1. | 24 1. | 24 1. | 24 1. | 24 2. |
| 25 1. | 25 1. | 25 2. | 25 1. | 25 1. | 25 2. | 25 2. |
|  |  | 26 2. |  |  | 26 2. |  |

**Fig. 15.** Some livelock scenarios in the program shown in Fig. 14, detected by DSCMC